# Metallization of the β-SiC(100) 3×2 Surface: a DFT Investigation

James Westover[1], Hamid Oughaddou[2,3], Hanna Enriquez[3], and Abdelkader Kara[1,2]

[1] Department of Physics, University of Central Florida, Orlando, FL 32816, USA
[2] Departement de Physique, Universite de Cergy-Pontoise, F-95031 Cergy-Pontoise Cedex, France
[3] *Département de Physique, Université Paris-Sud, Orsay Cedex 91405, France*

## Abstract

Using density functional theory (DFT) we report results for the electronic structure and vibrational dynamics of hydrogenated β reconstructed Silicon Carbide (001) (3x2) surfaces with various levels of hydrogenation. These results were obtained using density functional theory with a generalized gradient exchange correlation function. The calculations reveal that metallization can be achieved via hydrogen atoms occupying the second silicon layer. Further increases of hydrogen occupation on the second silicon layer sites result in a loss of this metallization. For the former scenario, where metallization occurs, we found a new vibrational mode at 1870 cm$^{-1}$, which is distinct from the mode associated with hydrogen atoms on the first layer. Furthermore, we found the diffusion barrier for a hydrogen atom to move from the second to the third silicon layer to be 258 meV.

# I. Introduction

Atomic hydrogen/deuterium (H/D) presents a single valence electron and is very well known to efficiently passivate semiconductor surfaces, both electrically and chemically [1-4]. On the other hand alkali metals, elements also from Group I, metallize semiconductors' surfaces and act as efficient promoters of semiconductor surface reactions such as oxidation or nitration [5,6]. Hence, it is noteworthy to mention the 1$^{st}$ example of H inducing metallization of a semiconductor surface has been shown for a β-SiC(100)3x2 reconstruction [7,8] along with an isotopic effect occurring when using deuterium [9]. Additionally, surface metallization induced by hydrogen adsorption has also been reported for ZnO and Ge [10,11].

Based on experimental (real-space atom-resolved scanning tunneling microscopy - STM, synchrotron radiation-based grazing incidence x-ray - GIXRD, and photoelectron diffraction - PED) and theoretical studies, the atomic structure of the Si-rich β-SiC(100) 3x2 reconstruction is now well known [8,12-15], and includes 3 Si atomic layers (1/3, 2/3 & 1 ML) on top of the 1$^{st}$ C atomic plane. The H/D atoms interaction with this reconstructed β-SiC(100) 3x2 surface has been investigated by different state-of-the-art experimental techniques including STM, scanning tunneling spectroscopy (STS) [7,8], valence band photoemission (VB-PES) [6-9], infrared absorption (IRAS) [7,16] and core level photoemission spectroscopies (CL-PES) [17]. These experimental investigations provided a model where H/D atoms are: i) decorating the top-most surface Si atoms' dangling bonds; ii) causing an asymmetric attack into the 3$^{rd}$ Si atomic plane (just above the 1$^{st}$ C plane). The latter was necessary to explain the observed extra component in the CL-PES data [17]. This Si-Si dimer was assumed to be broken upon adsorption of H/D on the 3$^{rd}$ plane, which would leave the H/D bonded to only one Si atom while the other Si atom would be left with a dangling bond. The fact that such a bond could only occur in the 3$^{rd}$ Si plane is further supported by CL-PES which shows a reactive component at the Si 2p core level [17]. This scenario would induce a significant charge transfer to the surface and especially to the 1$^{st}$ C-plane, leading to a large band bending [4, 17] comparable to those occurring during alkali metal/semiconductor interface formation [4]. Most interestingly, both charge transfer and band bending are significantly larger for the H-induced surface metallization compared to D-induced metallization clearly suggesting that such an isotopic effect is likely to result from a dynamical situation [9].

However, this model is *not* in full agreement with results obtained by *ab-initio* total energy calculations using density functional theory (DFT) approaches [18-23]. For nearly all of these calculations, there is agreement with the experimentally reported effects including: i) H-induced β-SiC(100)3x2 surface metallization, ii) H atoms decorating the topmost Si surface dangling bonds and iii) interacting in the 3$^{rd}$ Si atomic plane [18-21]. However, they all favor H atoms in a bridge bond position in the 3$^{rd}$ plane (hence the labeling "bridge-model") [18-21] instead of an asymmetric configuration proposed by four different types of experiments [7,8,16,17]. According to two of the DFT calculations, H atoms in a bridge bond position in the 3$^{rd}$ atomic plane between two Si atoms would induce vibrational modes around 1100 cm$^{-1}$ [20] and 1450 cm$^{-1}$ [21]. Such modes, however, were not observed in the IRAS experiments [6,12]. As mentioned above, the occurrence of an isotopic effect [9] suggests a dynamical situation with H/D atoms hopping from a Si dangling bond to another within the 3$^{rd}$ Si plane (Fig. 1a), especially at room temperature. One could therefore imagine that the bridge bonded H/D atom site becomes favored only at very low temperatures, as in the previous calculations [18-21], which were performed for a rigid system. Recently, two other theoretical investigations predicted H adsorption in the 2$^{nd}$ Si atomic plane, but with no noticeable metallization of the surface [22,23].

In this paper, we propose a new model, based on DFT calculations, where H atoms occupy the 2$^{nd}$ Si atomic layer in particular configurations that induce metallization of the surface. We will also present another configuration with H atoms on the 2$^{nd}$ Si atomic plane, which does not induce metallization. In the next section, we will present details of our calculations followed by a section containing results and discussion.

## II. Theoretical details

Electronic structure calculations have been performed using the Vienna *ab initio* simulation package (VASP) [24]. Exchange-correlation interactions were included within the generalized gradient approximation (GGA) in the Perdew-Burker-Ernzerhof form [25]. The electron-ion interaction is described by the projector augmented wave method as implemented by Kresse and Joubert [26]. Throughout the calculations we have used an energy cut-off of 250 eV and a special k-points mesh (5x4x1). The unit cell of our prototype system is shown in Fig. 1, where the surface is formed by a 1$^{st}$ silicon plane (whose density is 1/3 of that of a silicon plane in the bulk of SiC) containing 2 silicon atoms; followed by a 2$^{nd}$ silicon plane containing 4 silicon atoms and then 3 Si/C bilayers each containing 12 atoms; finally, the last silicon layer is passivated with 6 hydrogen atoms, while the top two silicon atoms are also passivated. The total number of atoms in our unit cell, before further adsorption of hydrogen atoms on the reconstructed surface, is 50 atoms. Every system has been relaxed until every force is less than 0.01 eV/Å.

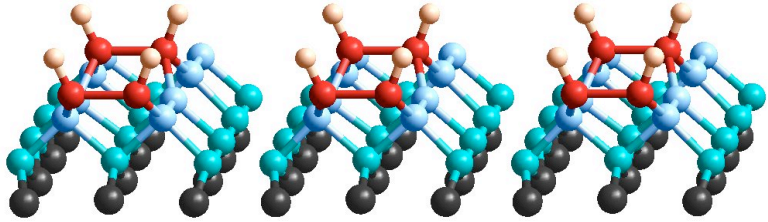

**Fig. 1:** Schematic of the β-SiC(100) 3x2 with hydrogen atoms

We have calculated the electronic and vibrational states of hydrogen and deuterium in a variety of situations. Only the relevant cases will be presented here. For the search of possible structures that form upon atomic hydrogen dosing we have adopted a different strategy than used in the previous theoretical investigations. After attaching two atoms to the first Si layer, we placed a hydrogen atom at about 1 Å above the surface with a random position in a plane parallel to the surface. When the system is allowed to relax, this third hydrogen atom is steered toward one of the four Si atoms in the second layer of our unit cell, independent of the initial position in the plane described above. For cases involving multiple H atoms in the second layer, it should be noted that the H atom was typically steered towards the nearest of the 4 Si atoms in the second layer in the relaxation process. This made it possible to predictably construct the configurations presented below.

## III. Results and discussion

Before we present our new results on the electronic structure and vibrational dynamics of hydrogen atoms decorating the second Si plane, let us reproduce the results of the previously studied "bridge-model". We started by decorating the top layer silicon atoms with hydrogen and subsequently added two hydrogen atoms forming a bridge between two silicon atoms in the third layer in the so-called "bridge model" configuration as shown in Fig.2a. The projected electronic density of states of Si (Fig. 2b) shows that the system changes from semi-conductor to metallic due to the appearance of a state crossing the Fermi level and centered at about 160 meV below it. Total energy calculations confirm this state to have the lowest energy configuration for H adsorption on β-SiC(100) 3x2 and to be characterized by vibrational modes at 1050 cm$^{-1}$ (polarized parallel to the surface) and at 1450 cm$^{-1}$ (polarized normal to the surface). Our results for the "bridge-model" are in good agreement with previous calculations. [20,21].

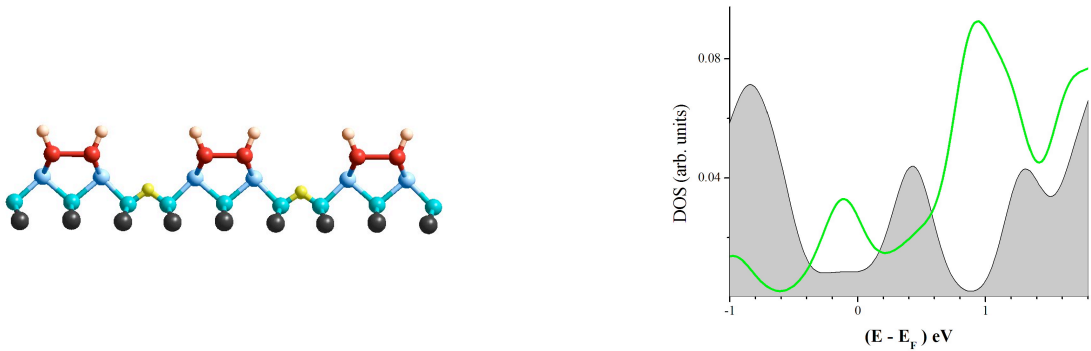

**Fig. 2:** a) side view of the bridge-model; and b) electronic density of states of the clean SiC surface (shade) and the bridge-model (line).

Using the structure in Fig. 1 as a starting configuration, we have launched an extra hydrogen atom as described in the previous section. This third hydrogen atom (the first two are decorating the first silicon layer) lands close to one of the silicon atoms in the second layer, as shown in Fig. 3a. This process has been repeated a few times with randomly chosen lateral position and the final position of the launched hydrogen atom was always close to one of the four silicon atoms on the second layer.

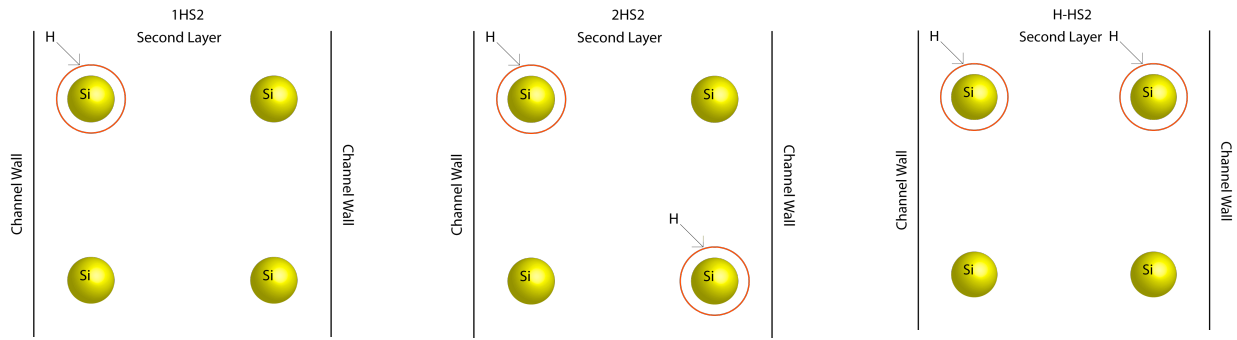

**Fig. 3:** Silicon atoms in the second layer decorated with a) one hydrogen atom; b) two hydrogen atoms on second-neighbors Si atoms; and c) two hydrogen atoms decorating neighboring silicon atoms.

A second hydrogen atom approaching the surface by the same manner also ended up decorating a second layer Si atom. The two hydrogen atoms may then be attached either to second nearest neighboring Si atoms (Fig. 3b) or to silicon nearest neighbors (Fig. 3c). We have calculated the electronic structure for these three configurations. In Fig. 4, we present the total density of states of surface silicon atoms in the three configurations.

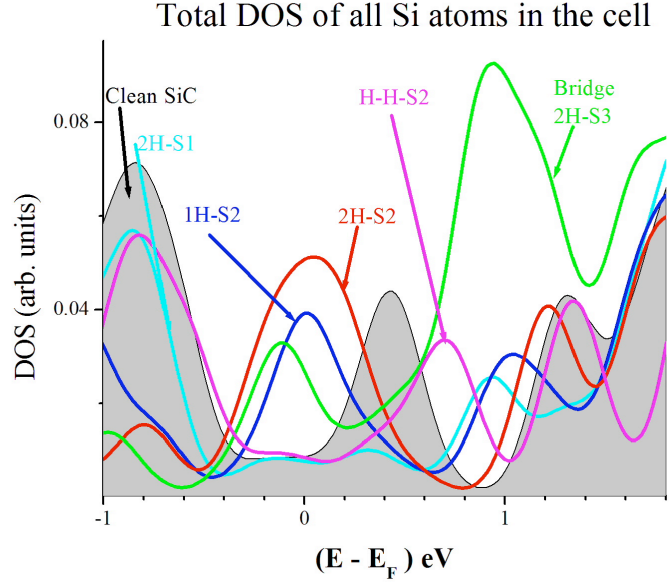

**Fig. 4:** Electronic densities of states for the studied systems. The labeling of 1H-S2, 2H-S2, and H-H-S2 correspond to the configurations in Fig 3a,b,c. The remaining configurations are Clean (no H on the surface), 2H-S1(H decorating only the uppermost Si atoms), and the bridge model.

From Fig. 4, it is clear that the two configurations in Fig. 3a and 3b induce metallization of the system. Unlike the previous configurations, the system in Fig. 3c maintains its original semi-conducting state. One may argue that the in the configuration 3b, the decorated silicon atoms are sufficiently far apart, making this configuration similar to that of Fig. 3a; that notion would depict what typically occurs at low hydrogen coverage. However, as the hydrogen coverage increases, neighboring silicon atoms start to be decorated, inducing the loss of metallization of the system. Hence, we conclude that as the hydrogen atoms approach the surface and after saturation of the first Si layer, they decorate the second Si layer atoms in a random fashion in the early stages of adsorption giving rise to metallicity. However, when the coverage becomes so large that hydrogen atoms start decorating neighboring silicon atoms, metallization starts to decrease.

This situation is metastable since the lowest energy state corresponds to having the additional H atoms in the bridge site between Si atoms in the third layer. In order for hydrogen atoms to reside in the second silicon layer, there should be a substantial diffusion energy barrier between hydrogen occupancy in the second layer and the third layer. We have thus calculated the diffusion barriers for a hydrogen atom to migrate from the second to the third layer and onto the fourth and fifth layers. The diffusion path was constructed from a one-dimensionally constrained path. A hydrogen atom attached to a silicon atom in the second layer is "dragged" [27] towards a silicon atom in the third layer in such a way that only the z-position of the hydrogen atom is changed and fixed during the diffusion. The z-position of the hydrogen atom was decreased by 0.1 Å until it reached the final configuration as depicted by the "bridge-model". At each stage, all (3N-1) degrees of freedom in the system were allowed to relax (the z-position of the hydrogen remains fixed). In Fig. 5, we show the total energy of the system vs. the z-position of the hydrogen atom. It is clear that a hydrogen atom decorating silicon atoms in the third layer is energetically more favorable than the configuration where it is decorating a silicon atom in the second layer. The barrier for a hydrogen atom to diffuse in this situation is found to 258 meV. This barrier may be higher if one includes the Van der Waals effects not considered in the present calculations. This value is not high enough to prevent a hydrogen atom from leaving its position in the Si second layer, however, it is high enough to keep the hydrogen atom in the second plane for some time. This scenario would lead to an accumulation of hydrogen atoms in the third layer, contrary to what is observed experimentally, since it is known that hydrogen can migrate towards the bulk of SiC [6]. It is not easy to calculate the migration of a hydrogen atom from third to fourth silicon layer as the path for diffusion is not easy to determine. A tentative calculation, using a 1D drag revealed a diffusion barrier larger than 1 eV, which can thus be possibly overcome by tunneling. Therefore, we conjuncture that a steady state is reached where hydrogen atoms residing in the second layer would diffuse to the third layer; and from there diffuse through tunneling to the bulk of SiC. Meanwhile, other atoms coming from the gas phase would occupy the vacant sites in the second layer, maintaining a random occupation of Si atoms in the second layer. Increasing the hydrogen flux would increase the coverage of the second layer and thus losing the metallization. This model is in accordance with the observation of H-Si-C vibrations using IRAS, a technique which is not particularly surface sensitive.

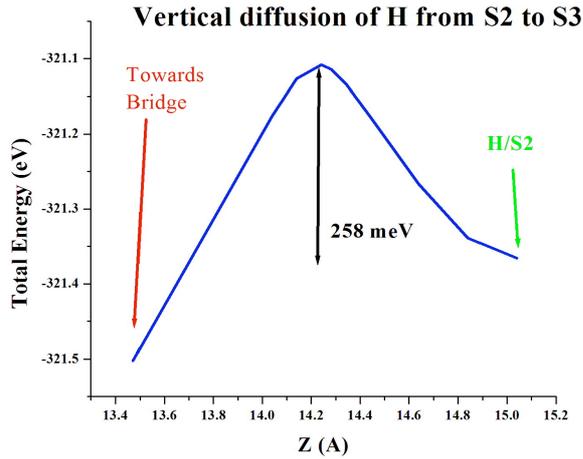

**Fig. 5:** Total energy of the system as a function of the z-position of the hydrogen atom

Now we turn to the vibrational dynamics of the system and in particular, the vibrational modes of hydrogen atoms, as these have been measured using different experimental techniques. Using finite differences and a parabolic fit, we have calculated the vibrational frequencies for hydrogen atoms occupying the second Si layer sites and found a mode whose value is 1870 cm$^{-1}$ with a polarization involving motion normal to the surface. This frequency shifts to 1322 cm$^{-1}$ for configurations using D. Note that this frequency is below the approximate 2000 cm$^{-1}$ value associated with vibrations of hydrogen atoms occupying the first layer.

## IV. Conclusions

In conclusion, based on the present investigation of β-SiC(100) 3x2 H (D) using DFT, we have proposed an alternative model for the hydrogen induced metallization of the surface. The investigation excludes that H may occupy the bridge site in the third Si layer given that the DFT predicted vibrational modes associated with the bridge model could not be observed experimentally. However, our DFT analysis shows that another model, based on the incomplete H decoration of the second Si layer is compatible with the existing experimental observations and predicts a new vibrational mode for H at 1870 cm$^{-1}$. Finally, we emphasize that our model for the H adsorption on SiC does not correspond to the lowest energy state correctly identified by DFT with H decorating the surface sites and occupying the bridge site between Si third layer atoms, but is indeed realized in nature because of the dynamical hindrance associated with the diffusion barriers present in this system.

**Acknowledgements:** AK thanks the university of Cergy Pontoise for hospitality. We thank E. Wimmer and P. Soukiassian for bringing the subject of this paper to our attention and for fruitful discussions. We also thank M. Rocca for fruitful discussions.

**References**

**1**-G.S. Higashi and Y.J. Chabal, in Handbook of Silicon Wafer Cleaning Technology: Science, Technology and Applications (ed. Kern,W.) 433–496 (Noyes, Park Ridge, New Jersey, 1993).

**2**-J.J. Boland, Phys. Rev. Lett. **65**, 3325 (1990); *ibid* **67**, 1539 (1991); Surf. Sci. **261**, 17 (1992).

**3**-K. Oura, V.G. Lifshits, A.A. Saranin, A.V. Zotov and M. Katayama, Surf. Sci. Rep. **35**, 1 (1999)

**4**-T. Seyller, J. Phys.: Condens. Matter **16**, S1755 (2004).

**5**-P. Soukiassian, M.H. Bakshi, H.I. Starnberg, Z. Hurych, T. Gentle, K. Schuette, Phys. Rev. Lett. **59**, 1488 (1987); P. Soukiassian, M.H. Bakshi, Z. Hurych, T.M. Gentle, Phys. Rev. B **35**, RC, 4176 (1987).

**6**-P. Soukiassian, M.H. Bakshi and Z. Hurych, J. Appl. Phys. **61**, 2679 (1987); P. Soukiassian, M.H. Bakshi, H. Starnberg, A.S. Bommannavar and Z. Hurych, Phys. Rev. B **37**, 6496 (1988).

**7**-V. Derycke, P. Soukiassian, F. Amy, Y.J. Chabal, M. D'angelo, H. Enriquez and M. Silly, Nature Mat. **2**, 253 (2003).

**8**-P. Soukiassian and H. Enriquez, J. Phys.: Condens. Matter **16**, S1611 (2004).

**9**-J. Roy, V.Yu. Aristov, C. Radtke, P. Jaffrennou, H. Enriquez, P. Soukiassian, P. Moras, C. Spezzani, C. Crotti and P. Perfetti, Appl. Phys. Lett. **89**, 042114 (2006).

**10**-Y. Wang, B. Meyer, X. Yin, M. Kunat, D. Langenberg, F. Traeger, A. Birkner and Ch. Wöll, Phys. Rev. Lett. **95**, 266104 (2005).

**11**-I.C. Razado, H.M. Zhang, G.V. Hansson and R.I.G. Uhrberg, Appl. Surf. Sci. **252**, 5300 (2006).

**12**-F. Semond, P. Soukiassian, A. Mayne, G. Dujardin, L. Douillard, C. Jaussaud, Phys. Rev. Lett. **77**, 2013 (1996).

**13**-M. D'angelo, H. Enriquez, V.Yu. Aristov, P. Soukiassian, G. Renaud, A. Barbier, M. Noblet, S. Chiang and F. Semond, Phys. Rev. B **68**, 165321/1-8 (2003).

**14**-A. Tejeda, D. Dunham, F.J. García de Abajo, J.D. Denlinger, E. Rotenberg, E.G. Michel and P. Soukiassian, Phys. Rev. B **70**, 045317/1-11 (2004).

**15**-W. Lu, P. Krüger, and J. Pollmann, Phys. Rev. B **60**, 2495 (1999).

**16**-F. Amy and Y.J. Chabal, J. Chem. Phys. **119**, 6201 (2003).

**17**-M. D'angelo, H. Enriquez, N. Rodriguez, V.Yu. Aristov, P. Soukiassian, A. Tejeda, E.G. Michel, M. Pedio, C. Ottaviani and P. Perfetti, J. Chem. Phys. **127**, 164716/1-10 (2007).

**18**-R. di Felice, C.M. Bertoni, C.A. Pignedoli and A. Catellani, Phys. Rev. Lett. **94**, 116103 (2005).

**19**-F. de Brito Mota, V.B. Nascimento and C.M.C. de Castilho, J. Phys.: Condens. Matter **17**, 4739 (2005); *ibid* **18**, 7505 (2006).

**20**-H. Chang, J. Wu, B.-L. Gu, F. Liu, and W. Duan, Phys. Rev. Lett. **95**, 196803 (2005).

**21**-X. Peng, P. Krüger and J. Pollmann, Phys. Rev. B **72**, 245320 (2005).

**22**-P. Deák, B. Aradi, J.M. Knaup and Th. Frauenheim, Phys. Rev. B **79**, 085314 (2009)

**23**-D.G. Trabada, F. Flores and J. Ortega, Phys. Rev B **80**, 075307 (2009)

**24-** G. Kresse and J. Furthmuller, Phys. Rev. B 54, 11169 (1996); G. Kresse and J. Furthmuller, Comput. Mater. Sci. 6, 15 (1996).G. Kresse and J. Hafner, Phys. Rev. B 47, 558 (1993).

**25-** J. P. Perdew, K. Burke, and M. Ernzerhof, Phys. Rev. Lett. 77, 3865 (1996).

**26-** G. Kresse and D. Joubert, Phys. Rev. B 59, 1758 (1999); P. E. Blochl, Phys. Rev. B 50, 17953 (1994).

**27-** H. Yildirim, A. Kara, S. Durukanoglu, and T.S. Rahman Surf. Sci. **600**, 484 (2006).